\documentclass[twocolumn]{article}

\usepackage{url}
\usepackage{graphicx}
\usepackage{subcaption}
\usepackage{color,colortbl}
\usepackage[utf8]{inputenc}
\usepackage[T1]{fontenc}
\usepackage{authblk}

\definecolor{Gray}{gray}{0.8}
\definecolor{LightGray}{gray}{0.9}

\title{Interdisciplinary Relationships Between Biological and Physical Sciences}

\author[1]{Paulo E. P. Burke}
\author[2]{Luciano da F. Costa}
\affil[1]{Bioinformatics Graduate Program, University of São Paulo, São Carlos, SP, Brazil}
\affil[2]{São Carlos Institute of Physics, University of São Paulo, PO Box 369, São Carlos, 13560-970, SP, Brazil}
\date{}

\begin{document}
	
	\maketitle
	
	\begin{abstract}
		Several interdisciplinary areas have appeared at the interface between biological and physical sciences. In this work, we suggest a complex network-based methodology for analyzing the interrelationships between some of these interdisciplinary areas, including Bioinformatics, Computational Biology, Biochemistry, among others. This approach has been applied over respective data derived from Wikipedia. Related reviews from the scientific literature are also considered as a reference, yielding a respective bipartite hypergraph which can be used to gain insights about the interrelationships underlying the considered interdisciplinary areas. Several interesting results are obtained, including greater interconnection between the considered interdisciplinary areas with biological than with physical sciences. A good agreement was also found between the network obtained from Wikipedia and the interrelationships revealed by the literature reviews. At the same time, the former network was found to exhibit more intricate relationships than in the hypergraph derived from the literature review.
	\end{abstract}
	
	\section{Introduction}
	
	Scientific research has become increasingly interdisciplinary ~\cite{Porter2009,VanNoorden2015}. It also has been suggested that interdisciplinary research tends to be intensely cited~\cite{Chen2015}.  Among all combinations of disciplines, the relationship between biological and physical sciences holds particular interest because of its potential impact on society.
	
	It is particularly challenging to trace back the origin of the interplay between biological and physical knowledge, perhaps as a consequence of the intrinsic relationship between these two areas, which tend to make almost any related research multidisciplinary. Nevertheless, we may consider as an important mark the application, in the 19th century, of chemistry techniques to the study of proteins\cite{Hunter2000}, which had been just discovered. The term ``Biochemistry'' was then coined in order to name the intersection of chemistry and biology.
	
	Another important starting-point that can be highlighted consists in the integration of mathematics and biology developed by Alfred J. Lotka and Vito Volterra's work on population dynamics in 1926, commonly known today as the predator-prey model~\cite{Berryman1992}. Decades later, the now well-established field of Biochemistry relied on the integration of concepts from several other areas. For example, in the early 60s computational approaches powered the elucidation of amino-acid sequences and three-dimensional structures of proteins\cite{Hagen2000}. At that moment, the most common term used to identify the application of mathematical and computational techniques to biological problems was ``Computational Biology''. Though still adopted today, this terminology has been almost subdued by the term ``Bioinformatics'', which was in great part motivated by the advent of DNA sequencing \cite{Sanger1977} in 1977, and further, to the Human Genome Project\cite{Watson1990} in 1990. Nowadays, some researches claim it is nearly impossible to achieve substantial scientific advances in biology without a strong computational basis\cite{Brodland2015,Markowetz2017}. In fact, the outcomes from such interactions have been central to answer fundamental questions of biology\cite{Han2008b,Nurse2011} and to push forward applied fields such as precision medicine\cite{Aronson2015,Krittanawong2017}.
	
	Given the fast growth and importance of the interface between biological and physical sciences through areas such as Biochemistry, Computational Biology, and Bioinformatics, it becomes important to understand how these areas interrelate one another. Though we could try understanding what is meant by each of these areas simplistically from their names (e.g., biochemistry as corresponding to chemistry in biology), the fast changes characterizing the development of modern science, and the appearance of new interfaces between branching subareas, makes a precise definition of the considered multidisciplinary areas very difficult.  Even so, efforts toward understanding, even in approximate terms, the meaning of existing multidisciplinary areas remain a critical issue for several reasons, including the need to continuously organizing knowledge and results, properly identifying interfaces and sub-areas, more effectively index the related literature, orient programs of study, among many other possibilities.
	
	Many are the works which aim at understanding how scientists manage interdisciplinarity~\cite{Palmer2002} and what is its impact on science~\cite{Omodei2017}. However, we lack approaches that can treat the organization of the interaction between different areas of knowledge \textit{per se}. Fortunately, due to the development of new information science concepts and methods, such as in complex networks~\cite{Silva2016a} and scientometry~\cite{Amancio2012,Silva2013,Mund2015}, it becomes possible to develop quantitative methods capable of revealing some of the main aspects of several inter and multidisciplinary areas through data. In this work, we propose a network approach to estimate knowledge interaction between scientific areas based on a given literature database. More specifically, we create a citation network based on articles related to areas of interest from a given database and their respective references. Then, we generate a higher-level network which captures the relationships between these areas based on the former citation network. As a case example, we use Wikipedia articles relative to biological and physical topics, as well as interdisciplinary areas which derive from them, in order to understand how the knowledge of these areas interact one another. We also perform a traditional literature review of the mentioned topics in order to have a reference for comparison.
	
	This paper is organized as follows. We first describe the general idea of constructing article networks and how we derive knowledge networks from it. Then, we present the Wikipedia database which is used to perform our analysis. The outcomes of applying our methodology to the Wikipedia database are then analyzed and discussed. At last, we present some conclusions about this work and possible future applications.
	
	\section{Methodology}
	
	Knowledge is mainly perpetuated through its register on some media that can be propagated and referred. Until now, the most efficient method found to store knowledge is to write it down in the form of books or articles, in paper or digitally. Also, it is very likely that most of the registered knowledge derives from, or relates to, some previous knowledge or information. In scientific literature, as well as in other areas, the relationship between pieces of literature are registered in the form of references between texts. Thus, the web formed by texts and references between them can unveil a higher-level structure of knowledge.
	
	In order to assess this web structure of knowledge, let us consider a collection of texts associated with a given subject. These texts, additionally to their contents, have references to other related texts, belonging or not to this collection. With articles and references, we can then construct an article network, following the procedure detailed in Section \ref{sec:network_construct}. Then, we can estimate the knowledge interactions between those subjects by grouping their respective articles. To avoid some unwanted biases, we normalize and group the nodes following the procedure described in Section \ref{sec:sampling}.
	
	\begin{figure*}[!ht]
		\centering
		\includegraphics[width=\linewidth]{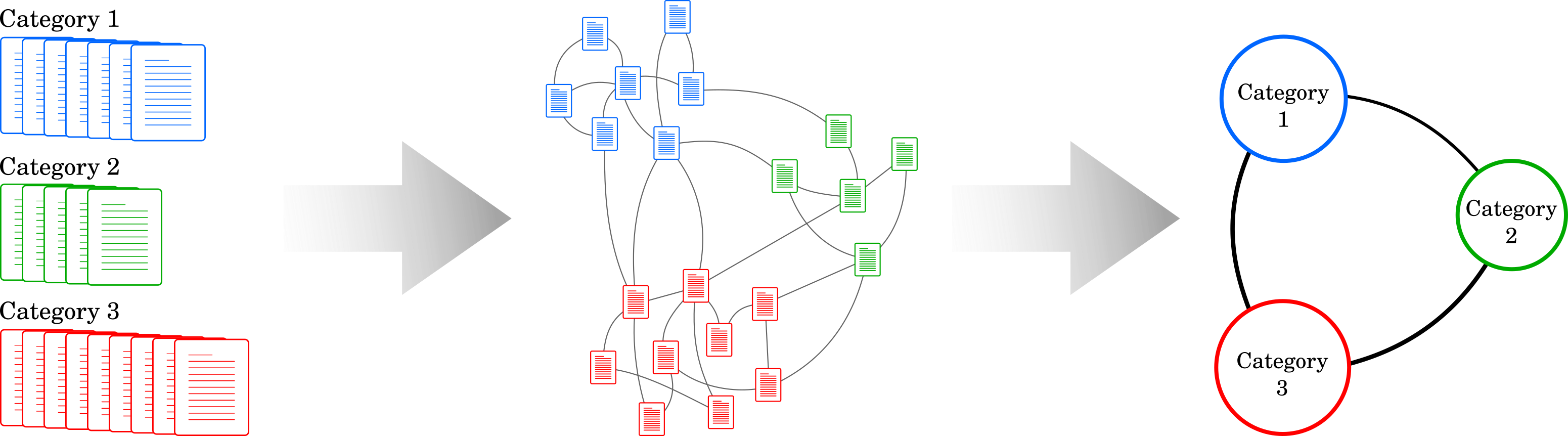}
		\caption{Illustration of the process of selecting articles, creating a citation network between them, and abstracting a respective knowledge network.}
		\label{fig:framework}
	\end{figure*}
	
	\subsection{Network Construction}\label{sec:network_construct}
	
	Networks have been used to model and analyze several kinds of data, from social interaction to transportation. Essentially, networks are composed of two elements: vertices and edges.  Real-world data involving discrete objects can be mapped into networks by assigning entities to vertices and using links to represent relationships between the entities. In the case of scientific literature, we can map texts into vertices and respective hyperlinks or bibliographic references as edges. In mathematical terms, our network can therefore be described as $G=(v,e)$, where $N$ is the total of texts, $v=\{v_1,v_2,\dots v_N\}$ are the vertices, and $e = \{(v_i,v_j),(v_k,v_l),\dots (v_m,v_n)\}$ are edges indicating respective relationships between the texts.
	
	\subsection{Normalization of Categories' Sizes by Sampling}\label{sec:sampling}
	
\begin{figure}[!ht]
		\centering
		\includegraphics[width=0.8\linewidth]{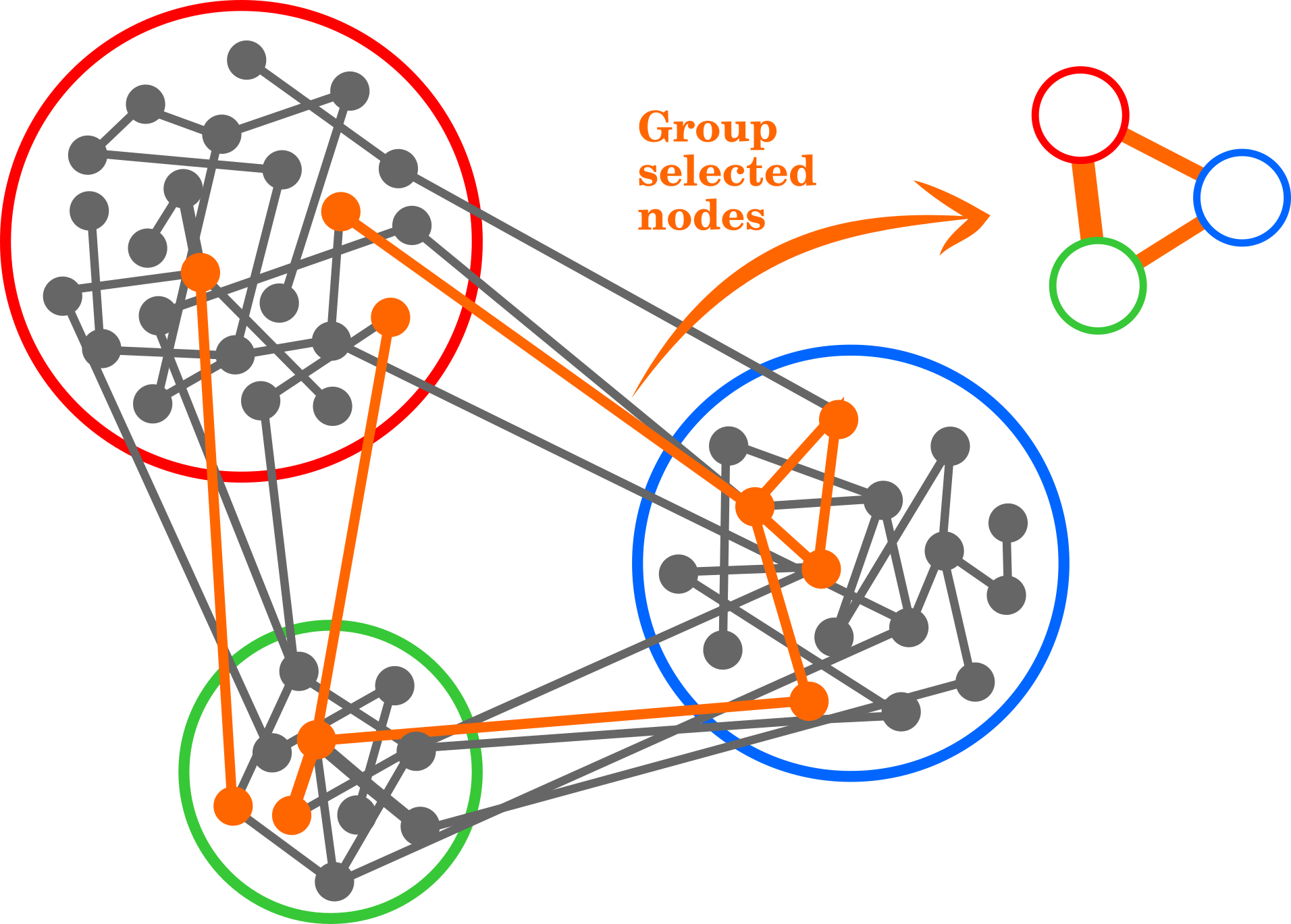}
		\caption{Illustration of the network sampling procedure. In this example, four nodes from each area are randomly selected while the links between them are kept. The resulting collapsed network has its links weighted by the total connections between areas yielded by the original nodes.}
		\label{fig:sampling}
	\end{figure}

	The number of articles contained in each of the categories of interest can substantially vary in magnitude. This variance can be a consequence of several factors, including area specificity, existence time, and funding, among others. This variation implies several difficulties while analyzing relationships between the considered areas.  For instance, the links between two larger areas would tend to appear in a more significant number than between two areas with a relatively smaller number of works. Thus, if one is interested in quantitatively comparing the relationship between any two different areas, it is necessary to normalize concerning the size of the areas. In order to minimize this potential bias, we sampled the network while selecting a fixed number of vertices for every one of its areas, as exemplified in Figure \ref{fig:sampling}. Observe that the number of selected vertices must be smaller than the size of the category containing the smallest number of articles.  Next, the articles so selected within each area were collapsed into a single vertex, resulting in a subsumed network where the number of vertices is equal to the number of areas. The frequency of connections between two areas, i.e.~the \emph{external edges}, is now represented as the weight of the edge which connects their respective vertices, while the frequency of connections within articles of the same area, the \emph{internal edges}, is stored as a node weight. After sampling the network several times, an average of the obtained networks will result in the normalized network of knowledge. Finally, the weight of each edge/node was divided by the sum of the weights of all edges/nodes in order to the result become invariant to the number of selected nodes per area.
	
	\subsection{Article Database}\label{sec:wiki}
	
	Wikipedia is currently the most extensive collaborative encyclopedia. Nowadays, the English version contains around 5.6 million articles with subjects ranging from science to popular culture, history, art, and others.
	
\begin{figure}[!htb]
		\includegraphics[width=\linewidth]{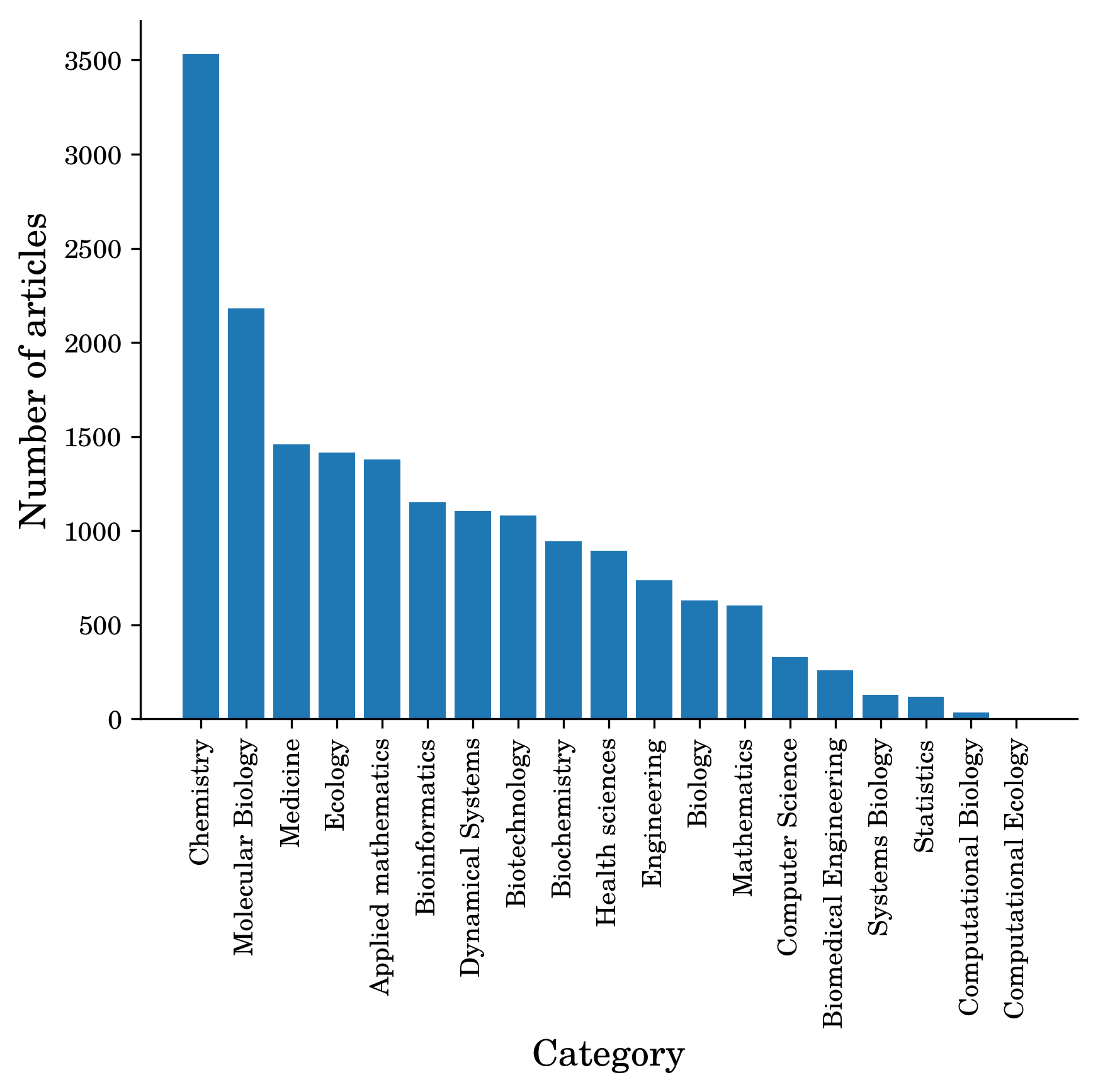}
		\caption{Number of articles in each of the considered categories.}
		\label{fig:num_articles}
	\end{figure}

	We selected a subset of articles from Wikipedia\footnote{The English version of the Wikipedia database can be freely downloaded from \url{dumps.wikimedia.org/enwiki}.} where their content refers to biological and physical sciences topics. More specifically, we selected articles originally assigned to the following categories: Chemistry, Mathematics, Applied mathematics, Dynamical Systems, Computer Science, Statistics, Engineering, Biomedical Engineering, Biology, Ecology, Medicine, Health Sciences, Molecular Biology, Bioinformatics, Biochemistry, Computational Ecology, Biotechnology, Systems Biology, and Computational Biology.   Observe that these categories correspond to respective tags available directly from Wikipedia.    Those categories were chosen in order to cover the main areas within biological and physical sciences as well as interdisciplinary areas, namely Bioinformatics, Computational Biology, Biomedical Engineering, Biochemistry, Computational Ecology and Systems Biology, which are intrinsically related to those two major areas. Given that Wikipedia's categories are hierarchically organized, we also selected subtopics at one hierarchical level downwards from the categories mentioned above. The total number of articles available in each category, considering their respective subtopics, is depicted in Fig. \ref{fig:num_articles}.
	
	\section{Results}
	
	Relationships between scientific areas can be estimated in several ways. Here, we used Wikipedia's articles as a means of quantitatively measuring the interactions between biological and physical areas, as well as interdisciplinary fields which might be a consequence of the interplay between these two major areas. In order to do so, we obtained articles from Wikipedia's database which were already assigned to labels referring to the scientific areas listed in Section \ref{sec:wiki}, as well as their respective sub-categories within one level in depth. However, some categories, namely Biomedical Engineering, Systems Biology, Statistics, Computational Biology, and Computational Ecology, were excluded from the analysis due to their very small number of articles when compared to the others categories, as observed in Figure \ref{fig:num_articles}.
	
\begin{figure*}[!ht]
		\centering
		\begin{subfigure}[b]{0.4\linewidth}
			\centering\includegraphics[width=\linewidth]{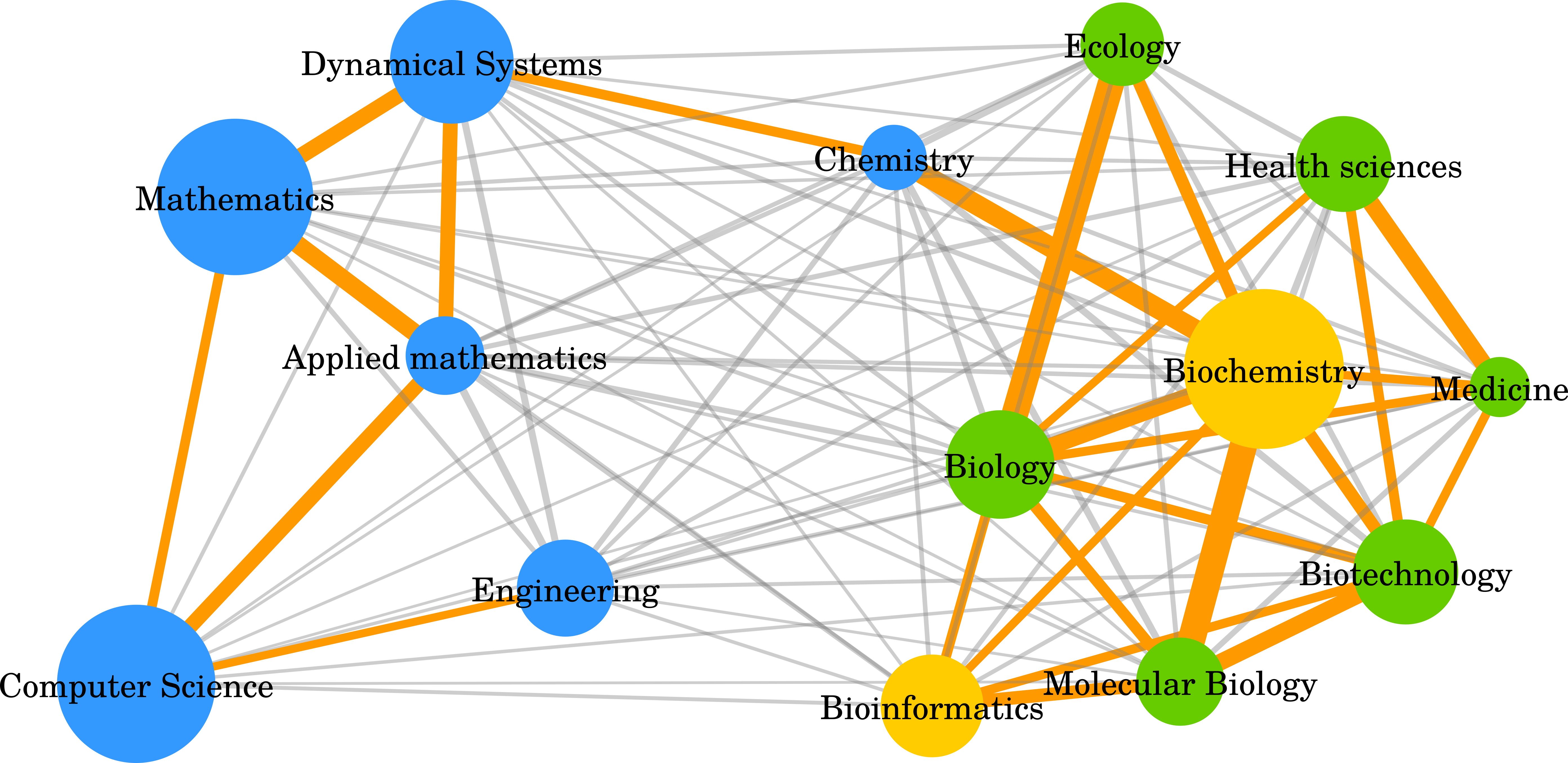}
			\caption{\label{fig:categorie_graph}}
		\end{subfigure}
		\begin{subfigure}[b]{0.25\linewidth}
			\centering\includegraphics[width=\linewidth]{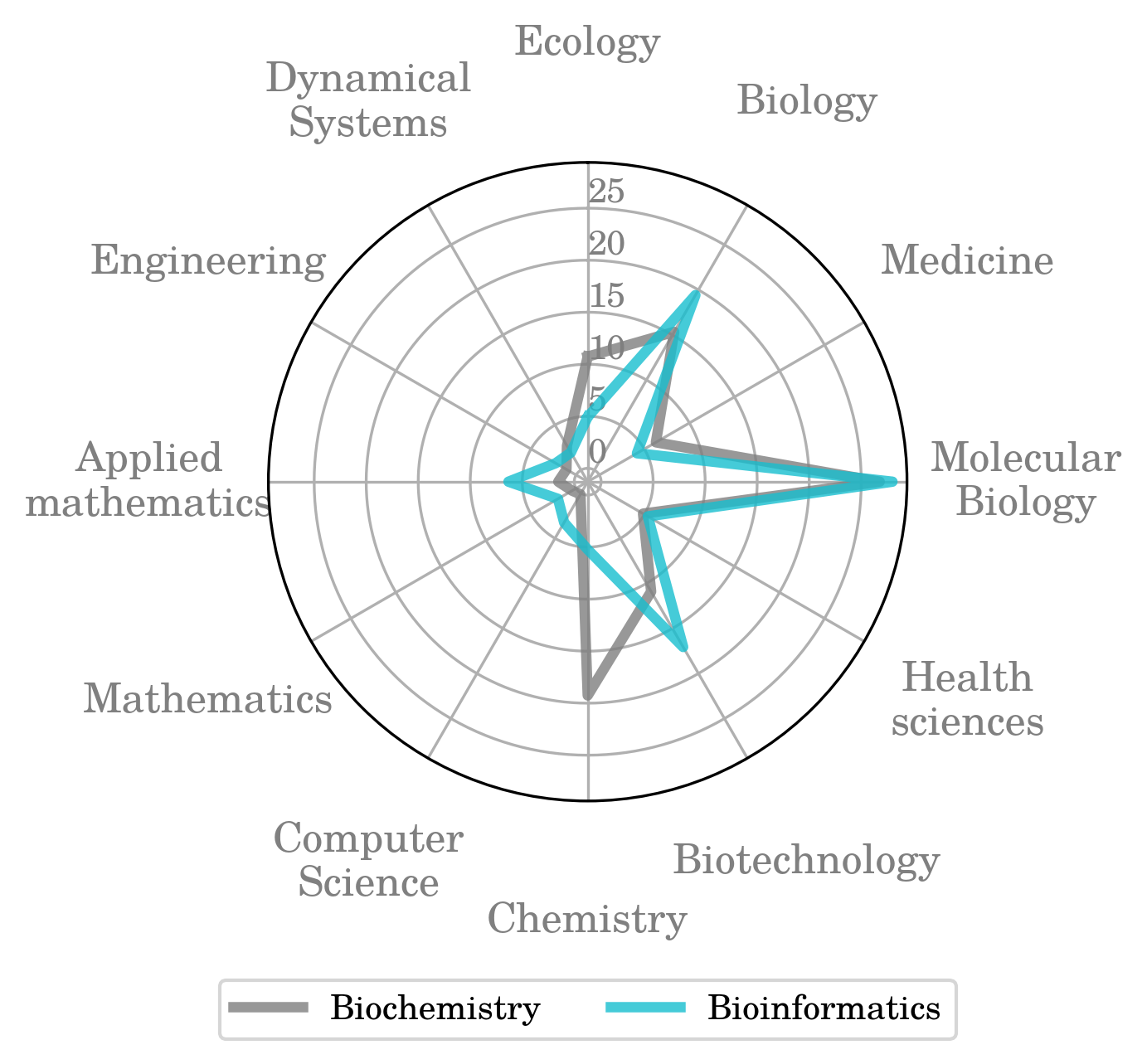}
			\caption{\label{fig:radar}}
		\end{subfigure}
		\begin{subfigure}[b]{0.32\linewidth}
			\centering\includegraphics[width=\linewidth]{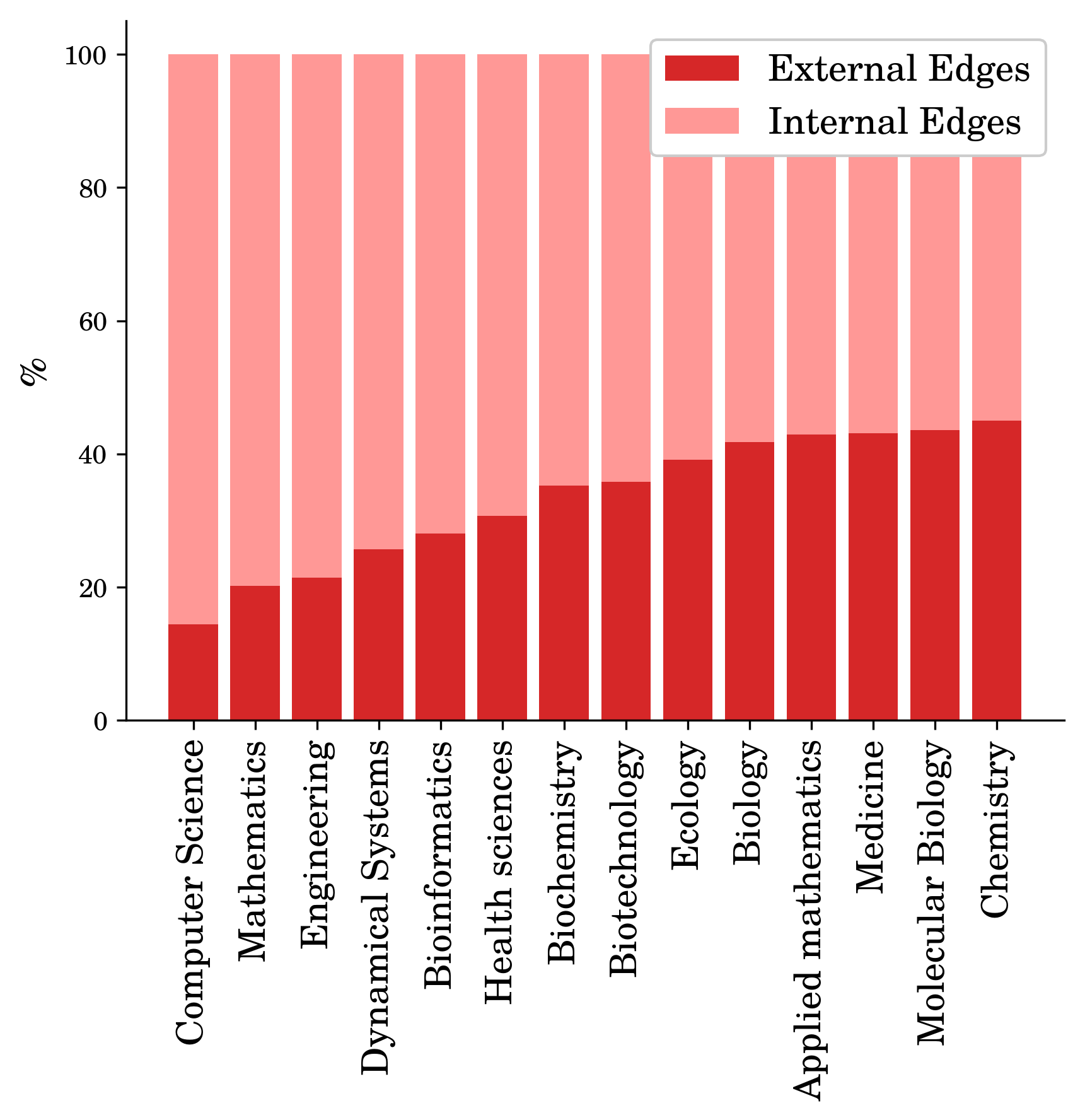}
			\caption{\label{fig:cat_InOut}}
		\end{subfigure}
		\caption{(a) Scientific areas network visualization. Biological, Physical, and Interdisciplinary areas are represented by green, blue, and yellow vertices, respectively. The size of each vertex is proportional to the number of connections within articles of the same area. The thickness of the edges is proportional to the number of links between the two areas it connects. Highlighted connections between areas compose the smallest set of strongest edges which keep all nodes connected into a single component. (b) Percentage of external connections of each interdisciplinary area with the other biological and physical areas. (c) The proportion of the internal and external edges of each area.}
	\end{figure*}

	We constructed the article network which contained 19,400 Wikipedia articles connected by 131,657 hyperlinks. To generate the knowledge network, we normalized the connections between areas following the procedure described in Section \ref{sec:sampling} by sampling 200 nodes of each area, and taking an average over 10,000 samples. We also sampled the network using 100 and 300 nodes per category, and no significant difference in the results was noticed (data not shown). The obtained knowledge network, depicted in Figure \ref{fig:categorie_graph}, is fully connected.  However, its strongest edges (highlighted in orange) show a well-bounded community structure forming two groups. The group on the left is composed solely of physical sciences, namely Mathematics, Applied Mathematics, Dynamical Systems, Computer Science and Engineering.
	
	On the other hand, the group on the right mixes all biological sciences, interdisciplinary areas, and Chemistry. This emergence of two groups indicates a substantial level of separation between physical and biological sciences. Interestingly, despite all interdisciplinary areas being strongly connected to biological sciences, they display different connecting patterns. In Figure \ref{fig:radar} we can observe that Biochemistry has most of its connections pointing to Chemistry, Molecular Biology, Ecology, and Biotechnology. However, even Bioinformatics also being strongly connected to the second group, in Figure \ref{fig:radar} we can observe that it has more connections with the physical group than the former area, being especially connected with Applied Mathematics and Computer Science.

	Regarding the balance between internal and external edges of each area, all fields have more internal than external edges. However, a trend can be observed in Figure \ref{fig:cat_InOut} where some areas have a higher percentage of external links than others. The first four areas with a lower percentage of external edges are all from physical sciences. Contrariwise, the last six areas which yield the highest external edges percentages comprise almost all biological areas plus Chemistry and Applied Mathematics. All interdisciplinary areas plus Health Sciences presented a midterm percentage of external edges in comparison to the others. This analysis suggests a more integrated environment within the biological than among physical sciences.

	\subsection{Interdisciplinary Relationships from Literature Review}
	
	In order to provide a comparative basis to our study, we gathered information about the disciplines that arose from the physical-biological interchange by performing a systematic literature review, looking for comprehensive texts which provide broad overviews or historical perspectives of their respective areas. We performed searches in Google Scholar using as keywords the areas' names combined with the words ``review'', ``overview'' and ``history''. Texts referring only to specific topics from the areas were not considered. The selected manuscripts were then organized in Table \ref{tab:literature} by row according to its interdisciplinary area and their reference in each column indicates that the article cites a relationship between the interdisciplinary area and the corresponding physical/biological field.
	
	\begin{table*}[]
		\caption{Literature review of interdisciplinary areas}
		\label{tab:literature}
		\begingroup
		\centering
		\renewcommand{\arraystretch}{1.5}
		\resizebox{\textwidth}{!}{%
			\begin{tabular}{|l|c|c|c|c|c|c|c|c|c|c|c|c|c|}
				\arrayrulecolor{white}
				& \multicolumn{6}{c}{\textbf{Physical Sciences}}                                                                                                                                                                                                                                                                                                                       & \multicolumn{6}{c}{\textbf{Biological Sciences}}                                                                                                                                                                                                                                                       &                                                                                        \\ \arrayrulecolor{Gray} \hline
				& \multicolumn{1}{c}{\textbf{Mathematics}} & \multicolumn{1}{c}{\textbf{\begin{tabular}[c]{@{}c@{}}Computer\\ Science\end{tabular}}} & \multicolumn{1}{c}{\textbf{\begin{tabular}[c]{@{}c@{}}Applied\\ Mathematics\end{tabular}}} & \multicolumn{1}{c}{\textbf{\begin{tabular}[c]{@{}c@{}}Dynamical\\ Systems\end{tabular}}} & \multicolumn{1}{c}{\textbf{Chemistry}} & \multicolumn{1}{c|}{\textbf{Engineering}} & \multicolumn{1}{c}{\textbf{Ecology}} & \multicolumn{1}{c}{\textbf{\begin{tabular}[c]{@{}c@{}}Molecular\\ Biology\end{tabular}}} & \multicolumn{1}{c}{\textbf{Biology}} & \multicolumn{1}{c}{\textbf{Biotechnology}} & \multicolumn{1}{c}{\textbf{Medicine}} & \multicolumn{1}{c|}{\textbf{\begin{tabular}[c]{@{}c@{}}Health\\ Sciences\end{tabular}}} &
				\multicolumn{1}{c|}{\textbf{Total}} \\ \hline
				\textbf{\begin{tabular}[c]{@{}l@{}}Computational\\ Ecology\end{tabular}}
				&              
				&   \cite{Helly1995,Pascual2005,Petrovskii2012}
				&   \cite{Helly1995,Pascual2005,Petrovskii2012}
				&   \cite{Helly1995,Pascual2005,Petrovskii2012}
				& 
				& 
				&   \cite{Helly1995,Pascual2005,Petrovskii2012}
				& 
				& 
				& 
				& 
				&
				&     3
				\\ \hline
				\rowcolor{LightGray}
				\textbf{Bioinformatics}
				&              
				&     \cite{Cohen2004,Luscombe2001,Stevens2013}
				&   \cite{Luscombe2001}
				& 
				& 
				& 
				& 
				&      \cite{Luscombe2001,Stevens2013}
				&   \cite{Cohen2004}
				& 
				& 
				& 
				& 3
				\\ \hline
				\textbf{\begin{tabular}[c]{@{}l@{}}Systems\\ Biology\end{tabular}}
				&          
				&   \cite{Hood2003,Kitano2002a,Kitano2002,Ideker2001,Kohl2010}
				&      \cite{Kitano2002a,Kitano2002,Kohl2010}
				& 
				& 
				& 
				& 
				&      \cite{Hood2003,Kitano2002a,Kitano2002,Ideker2001,Kohl2010}
				& 
				& 
				& 
				&
				& 5
				\\ \arrayrulecolor{Gray} \hline
				\rowcolor{LightGray}
				\textbf{\begin{tabular}[c]{@{}l@{}}Computational\\ Biology\end{tabular}}
				&   \cite{Waterman2000,Markowetz2017}           
				&   \cite{Slepchenko2002a,Waterman2000,Markowetz2017}
				&   \cite{Slepchenko2002a}
				&     \cite{Slepchenko2002a,Noble2002}
				& 
				& 
				& 
				&   \cite{Slepchenko2002a,Waterman2000,Noble2002}
				&   \cite{Markowetz2017,Noble2002}
				& 
				& 
				& 
				& 4
				\\ \arrayrulecolor{Gray} \hline
				\textbf{\begin{tabular}[c]{@{}l@{}}Biomedical\\ Engineering\end{tabular}}
				&   \cite{Bronzino2000,Enderle2012}
				&   \cite{Bronzino2000}
				& 
				& 
				& 
				&   \cite{Nebeker2002,Bronzino2000,Enderle2012}
				& 
				& 
				& 
				& 
				&      \cite{Nebeker2002,Bronzino2000,Enderle2012}
				&     \cite{Nebeker2002,Bronzino2000,Enderle2012}
				& 3
				\\ \hline
				\rowcolor{LightGray}
				\textbf{Biochemistry}
				&              
				& 
				& 
				& 
				&   \cite{Singh2004,Voet2010}
				& 
				& 
				&   \cite{Voet2010}
				&   \cite{Singh2004,Voet2010}
				& 
				& 
				&
				& 2
				\\ \hline
				\textbf{Total}
				&  4            
				&  15
				&  8
				&  5
				&  2
				&  3
				&  3
				&  11
				&  5
				&  0
				&  3
				&  3
				&
				\\ \hline
			\end{tabular}
		}
		\endgroup
	\end{table*}
	
	Regarding the considered biological and physical areas, we have from Table \ref{tab:literature}, Computer Science showed to be highly referred among almost all interdisciplinary sciences. It is intuitive to think that the vast majority of analyses that are carried out nowadays in any area make use of computational resources. On the biological side, Molecular Biology has a predominance in almost all interdisciplinary areas except Computational Ecology and Biomedical Engineering.
	
	With respect to the interdisciplinary areas, although we can say that Bioinformatics is today the most known name among all considered interdisciplinary areas, few were the scientific manuscripts found that discuss the area as a whole rather than some specific topic. Nevertheless, the so found manuscripts suggest an almost exclusive relationship between Computer Science and Molecular Biology. A similar scenario can be observed when considering Systems Biology. Despite being a younger scientific field, it underwent fast development at the beginning of this millennium due to its power to integrate biological data, and therefore, provide more faithful and larger computational models of biological systems.
	
\begin{figure}[!htb]
		\includegraphics[width=\linewidth]{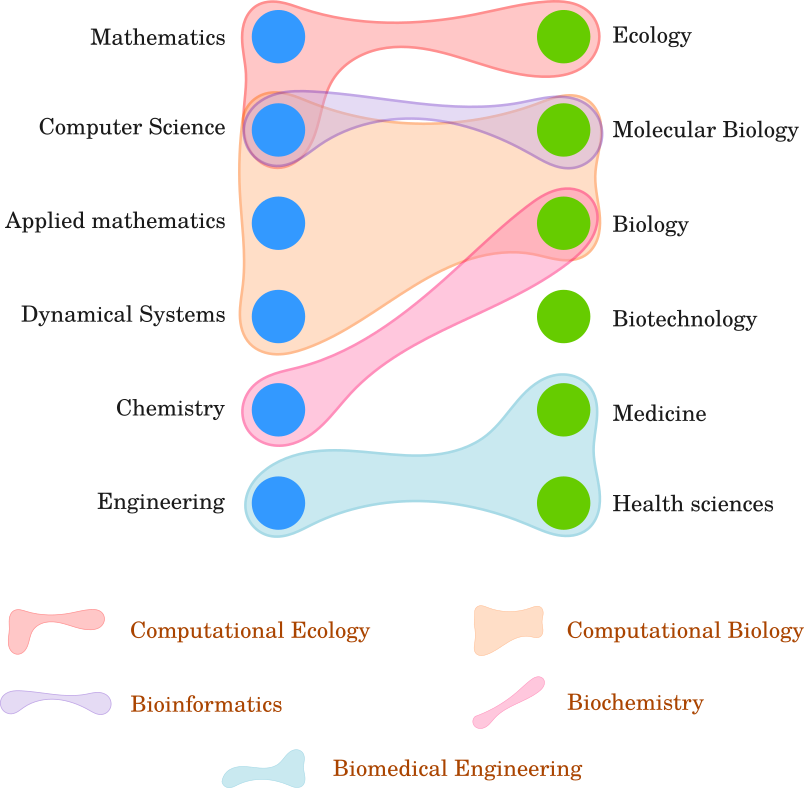}
		\caption{Diagram of relationships between biological and physical sciences derived from the  data in Table \ref{tab:literature}.}
		\label{fig:diagram}
	\end{figure}

	The data in Table \ref{tab:literature} is represented as a diagram in Figure \ref{fig:diagram}. It is easy to observe that Computational Biology extends through a wider range of areas, perhaps as a consequence of existing for a longer time. However, despite its historical importance, Biochemistry, our results suggest that it remained mostly within its original realm.
	
	If we compare the results obtained from Wikipedia with those derived from the performed literature review, some differences become evident. First of all, four out of all six considered interdisciplinary areas could not be considered in the Wikipedia analysis because of their extremely low number of articles.   The possible reasons for this relatively small number of include area specificity, existence time, and funding, among others.  On the other hand, for the remaining areas with many related articles in the Wikipedia analysis, namely Bioinformatics and Biochemistry, we could not find more than three and two respective review articles.
	
	It is worth noticing that the subject of Biotechnology, considered as a biological area, could not be found in any of the selected review articles. Nevertheless, we have from Figure \ref{fig:categorie_graph} that Biotechnology has strong interactions with both Bioinformatics and Biochemistry areas.

	\section{Concluding Remarks}
	
	As implied by its own name, \textit{interdisciplinary} sciences emerge from interplays between different disciplines, and their importance to scientific advance has grown steadily. Among all possible combinations between knowledge fields, the interaction between physical and biological sciences is one that deserves particular attention. The outcomes from this interaction are already affecting society as a whole, once they permeate vital sectors such as health and food production. Thus, it is essential to understand how the combinations of physical and biological sciences are establishing, evolving and influencing one another. To do so, we proposed a data-driven methodology where relationships between scientific areas can be estimated from the scientific literature. We used the Wikipedia's database to obtain articles whose categories were related to several scientific areas under biological and physical sciences, as well as interdisciplinary fields that have emerged from them.  Given that each of the considered areas had a substantially different number of articles, it was necessary to devise means for respective normalization, which was obtained by performing a sampling of the original data.   We also performed a literature review on the considered interdisciplinary topics in order to have a reference representing a more traditional approach.
	
	Several interesting results have been obtained.  For instance, we observed a possible greater approximation, in the sense of being more intensely interconnected, between the interdisciplinary areas with the biological instead of with the physical areas.  A possible explanation for this interesting structure could be that most of the interdisciplinary areas were originally motivated from biology, reflecting an increasing need for incorporating more and more concepts and tools from the physical sciences   In other words, the considered interdisciplinary areas would be mostly driven to biological applications.  Other interesting results include the centrality of Biochemistry among the biological areas, as well as the particularly strong interconnection between the considered biological and physical realms implemented by Chemistry.
	
	Regarding the investigation of review in the related scientific literature, we have that Computer Science and Molecular Biology are found in most of the reviews.  At the same, few reviews covered to a substantial extent the area of Biochemistry.  The table summarizing this literature review was then used to derive a bipartite hypergraph representation in which each interdisciplinary area is represented as respective hyperlink encompassing one or more areas from biological and physical sciences.  The so obtained hypergraph provides a putative and approximate characterization of the considered interdisciplinary areas.  At the same time, it is interesting to observe that, unlike this hypergraph, the network obtained from the Wikipedia data is completely interconnected.  This suggests that the interrelationships between the considered interdisciplinary areas could be more intricate and integrated than that highlighted by respective reviews.  Still, a good agreement can be observed between these two structures, with exception of the Biotechnology area, which appears well connected with other areas in the obtained network derived from Wikipedia while it was not highlighted in any of the considered reviews. 
	
	All in all, we have approached the problem of characterizing some of the main interdisciplinary areas between the biological and physical sciences in terms of both real data obtained from the Wikipedia, as well as through the consideration of some respective literature reviews.  Though a putative network of interrelationships has been obtained, it is important to bear in mind that these results are, as yet, substantially incomplete and preliminary.  The future availability of more data and literature reviews can provide a means for obtaining more complete and accurate characteristics by using the same, or similar, methodology.  Another interesting prospect for future research regards deriving descriptions of how the interdisciplinary areas changed with time.

	\section{Acknowledgments}
	
	Paulo E. P. Burke thanks Filipi Nascimento Silva for providing the filtered Wikipedia data. Luciano da F. Costa thanks CNPq (grant no.  307085/2018-0)  for  sponsorship. This  work  has benefited from FAPESP grant 15/22308-2. This study was financed in part by the Coordenação de Aperfeiçoamento de Pessoal de Nível Superior - Brasil (CAPES) - Finance Code 001.
	

\begin{thebibliography}{10}
		
		\bibitem{Porter2009}
		A.~L. Porter and I.~Rafols, ``{Is science becoming more interdisciplinary?
			Measuring and mapping six research fields over time},'' {\em Scientometrics},
		vol.~81, no.~3, pp.~719--745, 2009.
		
		\bibitem{VanNoorden2015}
		R.~{Van Noorden}, ``{Interdisciplinary research by the numbers},'' {\em
			Nature}, vol.~525, no.~7569, pp.~306--307, 2015.
		
		\bibitem{Chen2015}
		S.~Chen, C.~Arsenault, and V.~Larivi{\`{e}}re, ``{Are top-cited papers more
			interdisciplinary?},'' {\em Journal of Informetrics}, vol.~9, no.~4,
		pp.~1034--1046, 2015.
		
		\bibitem{Hunter2000}
		G.~K. Hunter, {\em {Vital forces : the discovery of the molecular basis of
				life}}.
		\newblock Academic Press, 2000.
		
		\bibitem{Berryman1992}
		A.~A. Berryman, ``{The Orgins and Evolution of Predator-Prey Theory},'' {\em
			Ecology}, vol.~73, pp.~1530--1535, oct 1992.
		
		\bibitem{Hagen2000}
		J.~B. Hagen, ``{The origins of bioinformatics.},'' {\em Nature Reviews
			Genetics}, vol.~1, no.~3, pp.~231--236, 2000.
		
		\bibitem{Sanger1977}
		F.~Sanger, S.~Nicklen, and A.~R. Coulson, ``{DNA sequencing with
			chain-terminating inhibitors},'' {\em Proceedings of the National Academy of
			Sciences}, vol.~74, pp.~5463--5467, dec 1977.
		
		\bibitem{Watson1990}
		J.~Watson, ``{The human genome project: past, present, and future},'' {\em
			Science}, vol.~248, no.~4951, pp.~44--49, 1990.
		
		\bibitem{Brodland2015}
		G.~W. Brodland, ``{How computational models can help unlock biological
			systems},'' {\em Seminars in Cell {\&} Developmental Biology}, vol.~47-48,
		pp.~62--73, dec 2015.
		
		\bibitem{Markowetz2017}
		F.~Markowetz, ``{All biology is computational biology},'' {\em PLOS Biology},
		vol.~15, p.~e2002050, mar 2017.
		
		\bibitem{Han2008b}
		J.~D.~J. Han, Y.~Liu, H.~Xue, K.~Xia, H.~Yu, S.~Zhu, Z.~Chen, W.~Zhang,
		Z.~Huang, C.~Jin, B.~Xian, J.~Li, L.~Hou, Y.~Han, C.~Niu, and T.~C. Alcon,
		``{Developmental systems biology flourishing on new technologies},'' {\em
			Journal of Genetics and Genomics}, vol.~35, no.~10, pp.~577--584, 2008.
		
		\bibitem{Nurse2011}
		P.~Nurse and J.~Hayles, ``{The cell in an era of systems biology},'' 2011.
		
		\bibitem{Aronson2015}
		S.~J. Aronson and H.~L. Rehm, ``{Building the foundation for genomics in
			precision medicine},'' 2015.
		
		\bibitem{Krittanawong2017}
		C.~Krittanawong, H.~J. Zhang, Z.~Wang, M.~Aydar, and T.~Kitai, ``{Artificial
			Intelligence in Precision Cardiovascular Medicine},'' 2017.
		
		\bibitem{Palmer2002}
		C.~L. Palmer, ``{Structures and strategies of interdisciplinary science},''
		{\em Journal of the American Society for Information Science}, vol.~50,
		no.~3, pp.~242--253, 2002.
		
		\bibitem{Omodei2017}
		E.~Omodei, M.~{De Domenico}, and A.~Arenas, ``{Evaluating the impact of
			interdisciplinary research: A multilayer network approach},'' {\em Network
			Science}, vol.~5, no.~2, pp.~235--246, 2017.
		
		\bibitem{Silva2016a}
		F.~N. Silva, D.~R. Amancio, M.~Bardosova, L.~d.~F. Costa, and O.~N. Oliveira,
		``{Using network science and text analytics to produce surveys in a
			scientific topic},'' {\em Journal of Informetrics}, vol.~10, no.~2,
		pp.~487--502, 2016.
		
		\bibitem{Amancio2012}
		D.~R. Amancio, M.~G. Nunes, O.~N. Oliveira, and L.~F. da~Costa, ``{Using
			complex networks concepts to assess approaches for citations in scientific
			papers},'' {\em Scientometrics}, vol.~91, no.~3, pp.~827--842, 2012.
		
		\bibitem{Silva2013}
		F.~N. Silva, F.~A. Rodrigues, O.~N. Oliveira, and L.~Da, ``{Quantifying the
			interdisciplinarity of scientific journals and fields},'' {\em Journal of
			Informetrics}, vol.~7, no.~2, pp.~469--477, 2013.
		
		\bibitem{Mund2015}
		C.~Mund and P.~Neuh{\"{a}}usler, ``{Towards an early-stage identification of
			emerging topics in science-The usability of bibliometric characteristics},''
		{\em Journal of Informetrics}, vol.~9, no.~4, pp.~1018--1033, 2015.
		
		\bibitem{Helly1995}
		J.~Helly, T.~Case, F.~Davis, S.~Levin, and W.~Michener, ``{The State of
			Computational Ecology},'' {\em San Diego Super Computer Center}, 1995.
		
		\bibitem{Pascual2005}
		M.~Pascual, ``{Computational Ecology: From the Complex to the Simple and
			Back},'' {\em PLoS Computational Biology}, vol.~1, no.~2, p.~e18, 2005.
		
		\bibitem{Petrovskii2012}
		S.~Petrovskii and N.~Petrovskaya, ``{Computational ecology as an emerging
			science.},'' {\em Interface focus}, vol.~2, pp.~241--54, apr 2012.
		
		\bibitem{Cohen2004}
		J.~Cohen, ``{Bioinformatics---an introduction for computer scientists},'' {\em
			ACM Computing Surveys}, vol.~36, pp.~122--158, jun 2004.
		
		\bibitem{Luscombe2001}
		N.~M. Luscombe, D.~Greenbaum, and M.~Gerstein, ``{What is bioinformatics? An
			introduction and overview},'' {\em Methods of Information in Medicine},
		pp.~346--358, 2001.
		
		\bibitem{Stevens2013}
		H.~Stevens, {\em {Life out of sequence : a data-driven history of
				bioinformatics}}.
		\newblock The University of Chicago Press, 2013.
		
		\bibitem{Hood2003}
		L.~Hood, ``{Systems biology: integrating technology, biology, and
			computation},'' {\em Mechanisms of Ageing and Development}, vol.~124,
		pp.~9--16, jan 2003.
		
		\bibitem{Kitano2002a}
		H.~Kitano, ``{Computational systems biology},'' {\em Nature}, vol.~420,
		pp.~206--210, nov 2002.
		
		\bibitem{Kitano2002}
		H.~Kitano, ``{Systems Biology: A Brief Overview},'' {\em Science}, vol.~295,
		pp.~1662--1664, mar 2002.
		
		\bibitem{Ideker2001}
		T.~Ideker, T.~Galitski, and L.~Hood, ``{A New Approach to Decoding Life :
			Systems Biology},'' {\em Annual Review of Genomics and Human Genetics},
		vol.~2, pp.~343--372, sep 2001.
		
		\bibitem{Kohl2010}
		P.~Kohl, E.~J. Crampin, T.~A. Quinn, and D.~Noble, ``{Systems Biology: An
			Approach},'' {\em Clinical Pharmacology {\&} Therapeutics}, vol.~88,
		pp.~25--33, jul 2010.
		
		\bibitem{Waterman2000}
		M.~S. Waterman, {\em {Introduction to computational biology : maps, sequences,
				and genomes : interdisciplinary statistics}}.
		\newblock Chapman {\&} Hall/CRC, 2000.
		
		\bibitem{Slepchenko2002a}
		B.~M. Slepchenko, J.~C. Schaff, J.~H. Carson, and L.~M. Loew, ``{Computational
			Cell Biology: Spatiotemporal Simulation of Cellular Events},'' {\em Annual
			Review of Biophysics and Biomolecular Structure}, vol.~31, pp.~423--441, jan
		2002.
		
		\bibitem{Noble2002}
		D.~Noble, ``{The rise of computational biology},'' {\em Nature Reviews
			Molecular Cell Biology}, vol.~3, no.~6, pp.~459--463, 2002.
		
		\bibitem{Bronzino2000}
		J.~D. Bronzino, {\em {The biomedical engineering handbook}}.
		\newblock CRC Press, 2000.
		
		\bibitem{Enderle2012}
		J.~D. J.~D. Enderle and J.~D. Bronzino, {\em {Introduction to biomedical
				engineering}}.
		\newblock Elsevier/Academic Press, 2012.
		
		\bibitem{Nebeker2002}
		F.~Nebeker, ``{Golden accomplishments in biomedical engineering},'' {\em IEEE
			Engineering in Medicine and Biology Magazine}, vol.~21, pp.~17--47, may 2002.
		
		\bibitem{Singh2004}
		P.~Singh, H.~S. Batra, and M.~Naithani, ``{History of biochemistry.},'' {\em
			Bulletin of the Indian Institute of History of Medicine (Hyderabad)},
		vol.~34, no.~1, pp.~75--86, 2004.
		
		\bibitem{Voet2010}
		D.~Voet and J.~G. Voet, {\em {Biochemistry}}.
		\newblock Wiley, 4~ed., 2010.
		
	\end{thebibliography}
	
	\bibliographystyle{ieeetr}
	
\end{document}